\documentclass[a4paper, amsfonts, amssymb, amsmath, reprint, showkeys, nofootinbib, twoside]{revtex4-1}

\usepackage{todonotes}
\usepackage{hyperref}
\hypersetup{colorlinks=true,
            linkcolor=blue,
            anchorcolor=blue,
            citecolor=blue}
\usepackage{setspace}
\begin{document}
%--------------------------------------------------------------------------
%  fill in the paper's title, author(s), and corresponding institutions
%--------------------------------------------------------------------------

\title{Dual-mode microresonators as straightforward access to \\octave-spanning dissipative Kerr solitons}

\author{{Haizhong Weng},$^{1}$ {Adnan Ali Afridi},$^{1}$ {Jing Li},$^{1}$ {Michael McDermott},$^{1}$ {Huilan Tu},$^{2}$ {Liam P. Barry},$^{3}$ {Qiaoyin Lu},$^{2}$ {Weihua Guo},$^{2,*}$ and {John F. Donegan}$^{1,}$\vspace{+2 mm}}
\email[Correspondence authors: ]{guow@hust.edu.cn}\email{jdonegan@tcd.ie}

\address{{$^{1}$}{School of Physics, CRANN and AMBER, Trinity College Dublin, Dublin 2, Ireland}\\{$^{2}$}{Wuhan National Laboratory for Optoelectronics and School of Optical and Electronic Information, Huazhong University of Science and Technology, Wuhan 430074, China}\\{$^{3}$}{Radio and Optical Communications Lab., School of Electronic Engineering, Dublin City University, Dublin 9, Ireland}}

\date{\today}

\begin{abstract}
\noindent
%---------------------------------------------------------------------------
%               Include abstract and keywords here
%---------------------------------------------------------------------------
The Kerr soliton frequency comb is a revolutionary compact ruler of coherent light that allows applications, from precision metrology to quantum information technology. The universal, reliable, and low-cost soliton microcomb source is key to these applications. In this work, we thoroughly present an innovative design strategy for realizing optical microresonators with two adjacent modes, separated by approximately 10 GHz, which stabilizes soliton formation without using additional auxiliary laser or RF components. We demonstrate the deterministic generation of the single-solitons that span 1.5-octaves, i.e., near 200 THz, via adiabatic pump wavelength tuning. The ultra-wide soliton existence ranges up to 17 GHz not only suggests the robustness of the system but will also extend the applications of soliton combs. Moreover, the proposed scheme is found to easily give rise to multi-solitons as well as the soliton crystals featuring enhanced repetition rate (2 and 3 THz) and conversion efficiency greater than 10$\%$. We also show the effective thermal tuning of mode separation for stably accessing single-soliton. Our results are crucial for the chip-scale self-referenced frequency combs with a simplified configuration.

\end{abstract}
\maketitle
%---------------------------------------------------------------------------
%               the main text of your paper begins here
%---------------------------------------------------------------------------

\section{Introduction}\label{sec:intro}
\noindent Dissipative Kerr soliton (DKS), as a self-reinforcing wave packet that maintains its shape while circulating around a microresonator, has been demonstrated under a double balance between nonlinearity and dispersion, as well as parametric gain and cavity loss \cite{doi:10.1126/science.aad4811, kippenberg2018dissipative}. Due to the unprecedented compactness, low-noise, high power-efficiency, and broad spectral bandwidth, soliton Kerr combs (microcombs) have attracted considerable research interest and been extensively studied for spectroscopy \cite{doi:10.1126/science.aah6516}, communications \cite{marin2017microresonator}, frequency synthesizer \cite{spencer2018optical}, optical clock \cite{drake2019terahertz}, microwave photonics \cite{liu2020photonic} and sensor applications \cite{yao2021}. Over the past several years, through the substantial exploration of the fundamental physics and microresonator fabrication, researchers have realized Kerr solitons in a growing number of platforms, including ultra-high \textit{Q} MgF$_2$ \cite{herr2014temporal}, silica \cite{yi2015soliton}, and monolithic integrated platforms such as Si$_3$N$_4$ \cite{doi:10.1126/science.aad4811, joshi2016thermally, wang2016intracavity, ye2021integrated}, LiNbO$_3$ \cite{he2019self}, AlGaAs \cite{moille2020dissipative} and Ta$_2$O$_5$ \cite{jung2021tantala}, as well as the wide-bandgap semiconductors AlN \cite{weng2021directly, liu2021aluminum}, SiC \cite{guidry2021quantum} and GaN\cite{https://doi.org/10.1002/lpor.202100071}. 

 Photonic integration of laser pump and passive resonators offers the possibility of achieving chip-scale operation, but there are significant challenges to overcome before the widespread deployment of these soliton comb systems. One key challenge comes from the thermo-optic instability in the microresonator when the pump enters into the red-detuned regime for soliton formation. To stably access the soliton state, a number of experimental techniques were developed including rapid laser frequency scanning \cite{herr2014temporal, liu2021aluminum, jung2021tantala}, careful pump power manipulation \cite{doi:10.1126/science.aad4811, yi2015soliton}, or microheater thermal tuning \cite{joshi2016thermally, ji2021exploiting}. With the extra radio-frequency (RF) generator, modulator, or microheaters, these schemes can bring the short-lived soliton to a steady state. Self-injection locking (SIL) \cite{stern2018battery, raja2019electrically} has been proposed and exploited for turnkey soliton generation \cite{shen2020integrated} or even a remarkable octave-spanning soliton microcomb \cite{briles2021hybrid}, by directly coupling a laser chip to a passive microresonator. This approach enables a miniaturized frequency comb source but demands challenging photonic integration and great caution to control the back-reflected Rayleigh scattering. Also, the likely accessible detuning reduction in SIL will limit the spectral bandwidth and dispersive waves (DWs) intensity \cite {briles2021hybrid}, as well as the SER and total comb power \cite{voloshin2021dynamics}, which are serious issues for the precision metrology and timing. Recently, dual-pumping for two resonances (2P2R) has been applied to mitigate the thermal effects thus leading to the deterministic generation and switching of the DKS \cite{https://doi.org/10.1002/lpor.202100071, zhang2019sub, zhou2019soliton, lu2019deterministic}. However, it drastically increases the system complexity and cost due to the use of another set of laser, amplifier, polarization controller, and fiber circulator \cite{https://doi.org/10.1002/lpor.202100071, zhou2019soliton, lu2019deterministic}. Another dual-pumping scheme but activating a single resonance (2P1R) was also proposed, where the auxiliary pump is one modulation sideband away from the main pump \cite{wildi2019thermally, nishimoto2022thermal}. 

\begin{figure}[tb]
\centering
\includegraphics[width=1\linewidth]{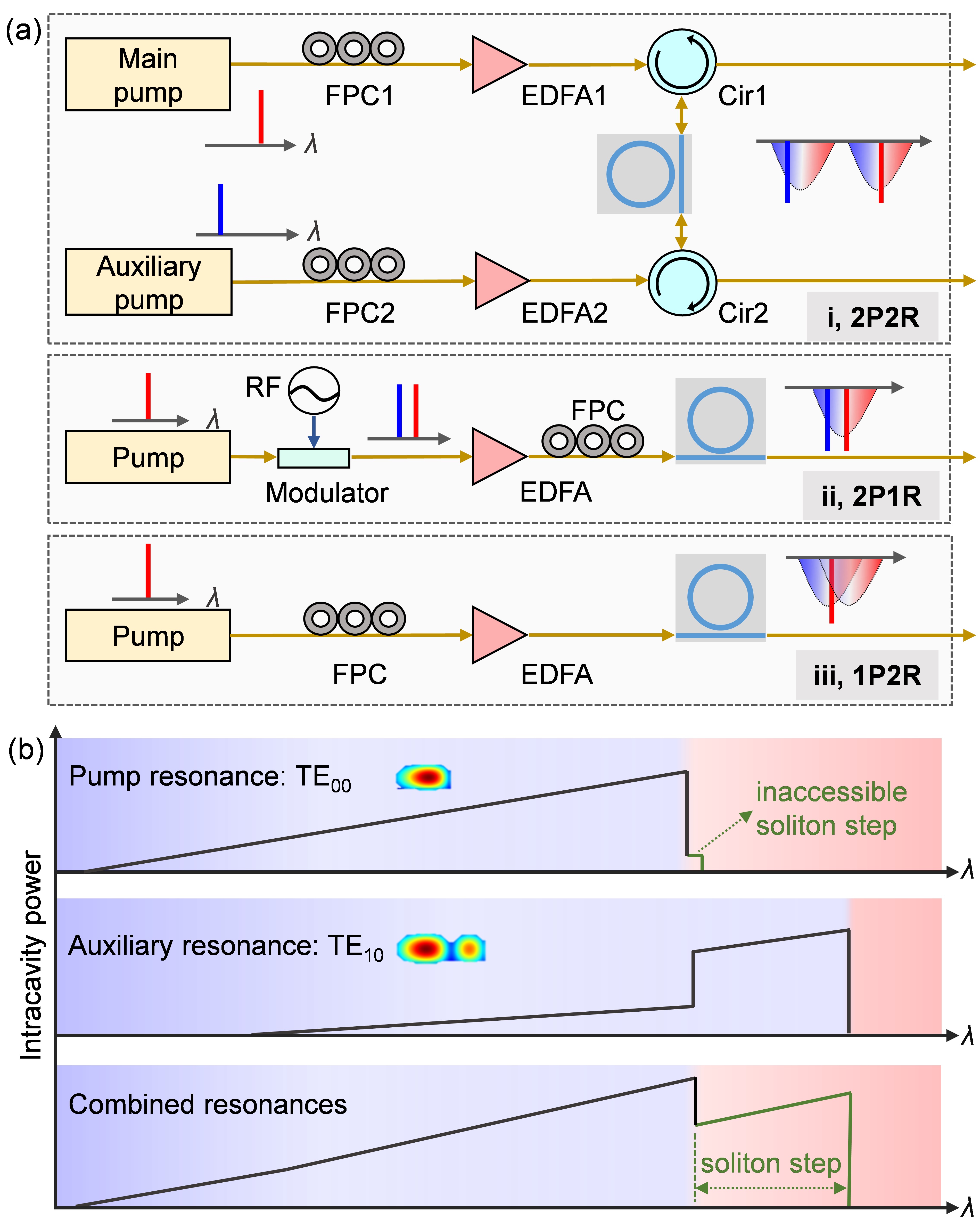}
\caption{\textbf{(a)} Schematic of different thermal compensation schemes for Kerr soliton generation. FPC: fiber polarization controller; EDFA: erbium-doped fiber amplifier; Cir: circulator. \textbf{(b)} Intracavity power versus the pump detuning for 1P2R system. The power coupled into the auxiliary mode rises suddenly due to the blue shift of the resonances during soliton formation, which can, in turn, remedy the intracavity power change thus preventing the soliton collapse. Blue and red shaded indicate the pump detuning position relative to the cavity resonance.}
\label{fig1}
\end{figure}

A simple and cost-effective octave-spanning Kerr soliton generator needs to be developed urgently to bring the microcombs towards practical applications. Here, we present the straightforward access to octave-spanning DKSs by injecting a single pump to two close resonances (dual-mode) with the same polarization, which we called the 1P2R-1P scheme. The modes on the blue and red sides are used for parametric processes and thermal compensation, respectively. Figure \ref{fig1} compares different thermal accessible soliton systems and sketches the 1P2R mechanism. In contrast to dual-pumping, our method leverages a much simplified setup once the front-end design is properly managed. The idea used in this paper was proposed by our group and successfully applied for octave-spanning DKS generation in an AlN microring resonator (MRR) \cite{weng2021directly}. Dual-mode but with mixed polarization (1P2R-2P) was also found to help the soliton stabilization process \cite{li2017stably}, which requires careful adjustment of the polarization. However, the real potential of the 1P2R scheme needs investigation due to the rigorous requirements in design and fabrication for ensuring the pump and auxiliary modes are in close proximity.

This paper presents the production of octave-spanning Kerr solitons with improved performance through careful design of the microresonators while we also discuss the feasibility of fabrication with high yield. The advanced single-soliton features a 17-GHz-wide soliton existence range (SER) and a 200-THz-wide spectral bandwidth. SER means an effective detuning range where a soliton state is maintained during the pump wavelength tuning \cite{herr2014temporal}. Moreover, using the same resonance, octave-spanning soliton crystals at the telecommunication C-band are also demonstrated. Similar soliton behavior are also observed in multiple chips, and thereby illustrate its universal nature. The presented results provide a solid strategy for broadband DKS generation, which is transferable to alternative materials with a tailored repetition rate (\textit{f}$_{rep}$). From an application perspective, the 1P2R-1P scheme paves the way towards making reliable, dynamic, low-cost, and easy-to-operate soliton microcomb sources.

\begin{figure*}[t]
\centering
\includegraphics[width=1\linewidth]{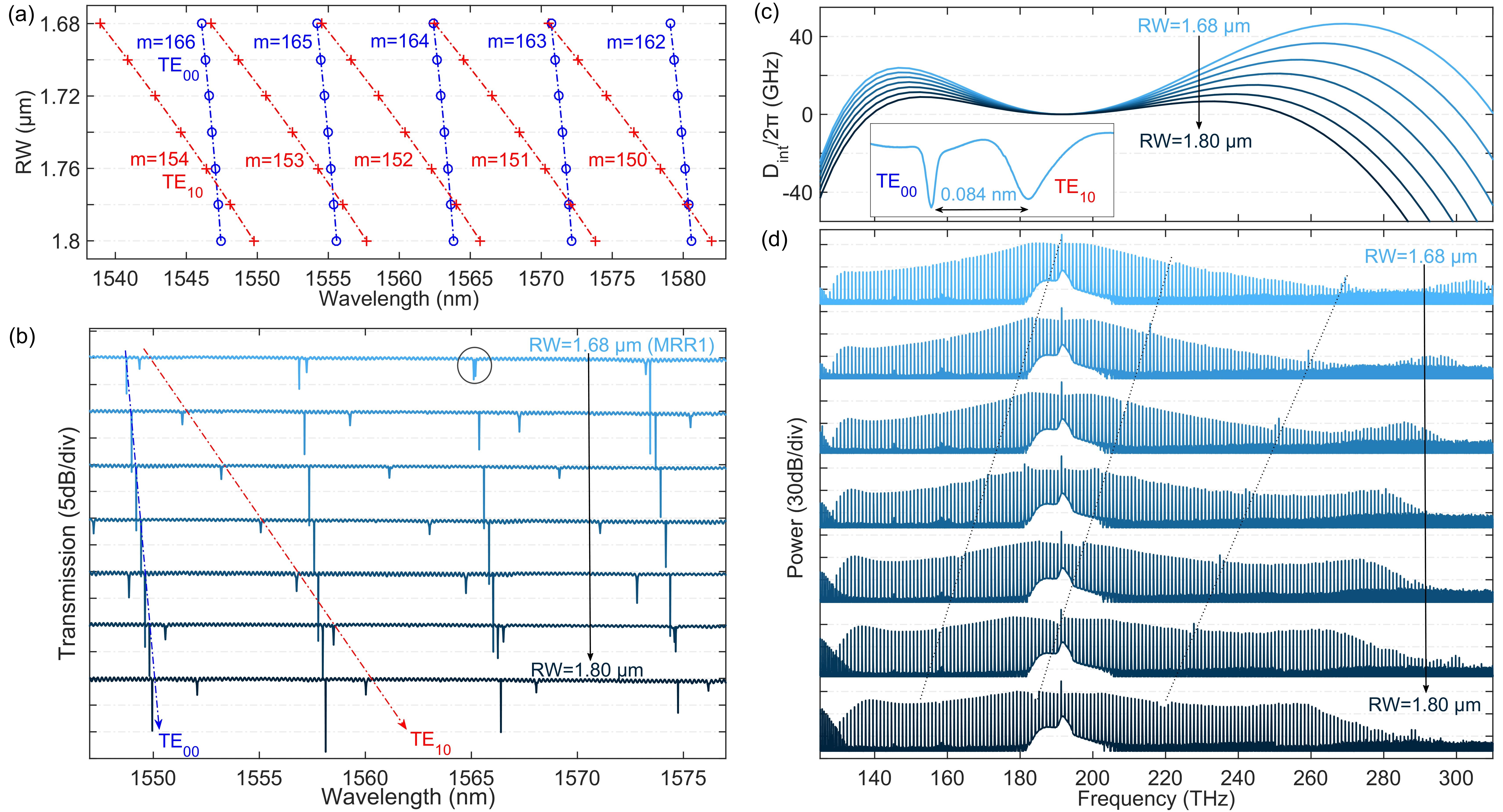}
\caption{\textbf{(a)} Simulated resonant wavelengths for the microrings with a radius of 23.25 $\mu$m, the thickness of 800 nm, and various RWs. \textbf{(b)} Measured transmission spectra of seven adjacent MRRs with various RWs. RW change step is 20 nm. The circle denotes the target dual-mode enabling soliton generation with the 1P2R-1P scheme. \textbf{(c)} Simulated integrated dispersion profiles. Inset: zoom-in view of the dual-mode. \textbf{(d)} Measured MI comb spectra at a 400 mW on-chip power for varying RW. Dashed lines indicate the mode interaction positions.}
\label{fig2}
\end{figure*}

\begin{figure*}[t]
\centering
\includegraphics[width=1\linewidth]{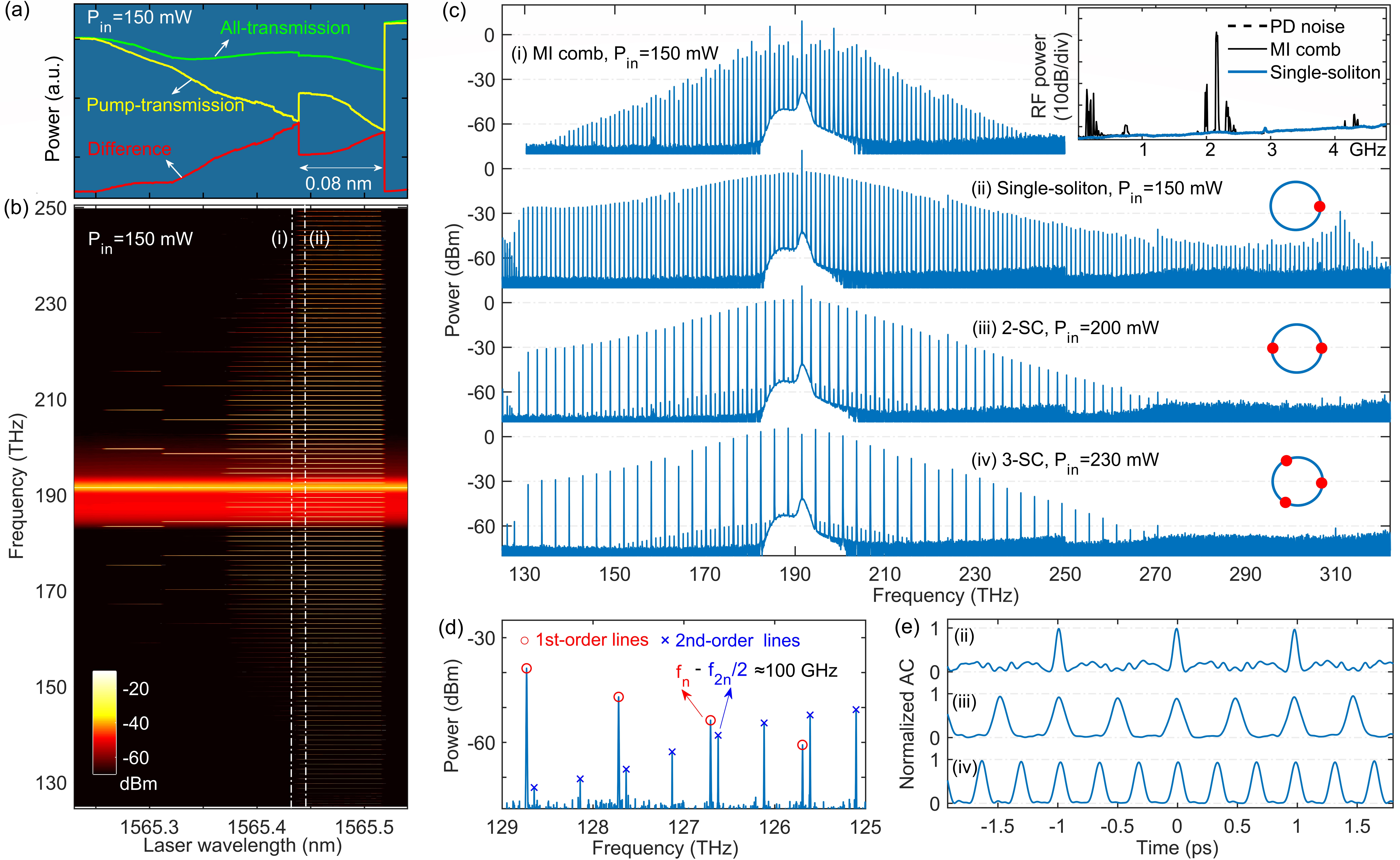}
\caption{Experimental soliton results of MRR1. \textbf{(a)} Collected powers of all output (green) and the pump alone (yellow), as well as their difference (red) at \textit{P}$_{in}$=150 mW. \textbf{(b)} Microcomb evolution map. \textbf{(c)} Optical spectra of (i) MI comb, (ii) single-soliton, (iii) 2-SC and (iv) 3-SC by connecting output fiber with the OSA directly. Top inset: RF noise of MI comb and single-soliton. The photodiode (PD) noise floor is overlapped by that of the single-soliton. Lower insets describe the solitons distribution. \textbf{(d)} Single-soliton spectrum at low-frequency. \textbf{(e)} Measured autocorrelation (AC) traces of various soliton states.}
\label{fig3}
\end{figure*}

\section{Design of dispersion-engineered dual-mode microresonators}\label{sec:design}
Figure \ref{fig2} shows the design and resonance characteristics of the dual-mode MRRs. The 800-nm-thick Si$_3$N$_4$ MRRs that we are using were fabricated by Ligentec through photonic Damascene processes \cite{liu2021high}. To build a reliable layout design, we first conducted the simulations with finite element method. For the MRRs with a fixed radius of 23.25 $\mu$m and varied ring widths (RWs), the simulated resonant wavelengths are plotted in Fig. \ref{fig2}(a). As the RW increases, an obvious mode redshift is observed, while the slope d$\lambda$/d\textit{RW} is 0.012 and 0.094 for the TE$_{00}$ and TE$_{10}$ modes. Consequently, a 1 nm RW variation will lead to a $\sim$0.08 nm adjustment in the mode separation $\Delta\lambda$ ($\lambda_{10}$-$\lambda_{00}$), i.e., a 100 nm RW variation will lead to a $\sim$8 nm $\Delta\lambda$ change, which is almost one free spectra range (FSR). Specifically, the TE$_{00}$ mode with an angular number (\textit{m}) of 164 and TE$_{10}$ mode (\textit{m}=151) have a minimum $\Delta\lambda$ of 0.09 nm at $\sim$1563 nm when RW=1.68 $\mu$m. Then the two modes coincide again at $\sim$1572 nm with a separation of 0.12 nm when RW=1.78 $\mu$m. 

Figure \ref{fig2}(b) displays the measured transmission spectra of seven neighbouring MRRs with RWs discretely increasing from 1.68 to 1.80 $\mu$m. As the RW increases, both modes show the redshift (denoted by arrows) with the speeds almost the same as simulations, illustrating that a 20 nm variation in the MRR dimension is achievable even using the 248 nm DUV lithography. The target dual-mode in MRR1 (denoted by a circle) has a $\Delta\lambda$ of 0.084 nm, i.e., $\sim$10.5 GHz, and is enlarged as an inset of Fig. \ref{fig2}(c). The TE$_{00}$ and TE$_{10}$ modes have an FSR of $\sim$1012 and $\sim$979 GHz, respectively. Their statistical \textit{Q} factors are 1.1×10$^6$ and 4.2×10$^5$, respectively. The dual-mode position shifts to $\sim$1566.3 nm when RW=1.78 $\mu$m, while the microcombs cannot be tuned to the soliton regime due to a relatively large $\Delta\lambda$ of 0.262 nm. Overall, the experimental results are consistent with simulations, suggesting that dual-mode resonators in the C-band can be reliably attained by tailoring the MRR dimensions. 

Near-zero anomalous dispersion is crucial for broadband microcomb generation. Figure \ref{fig2}(c) presents the calculated integrated dispersion (D$_{int}$) \cite{doi:10.1126/science.aad4811} of the TE$_{00}$ mode family. All the MRRs can support dual-DW, where the D$_{int}$ equals zero. With the increase of RW, the DW position at low-frequency has a small blue shift, while the high-frequency DW dramatically drops from 311 to 254 THz. The simulated second-order dispersion D$_{2}$/2$\pi$ is 37.2 and 19 MHz when RW is 1.68 and 1.78 $\mu$m, respectively. Figure \ref{fig2}(d) summarizes the measured modulation instability (MI) microcombs when pumping the MRRs with an on-chip power of \textit{P}$_{in}$=400 mW. All spectra exceed an octave span thanks to the dual-DW. The groups of dense lines below 130 THz result from the 2nd-order diffraction of the optical spectrum analyzer (OSA), and are thus artifacts. The peculiar comb lines with enhanced or reduced power (denoted by dotted lines) are caused by mode interactions \cite{ramelow2014strong}. The wider MRRs tend to have flatter spectral envelopes and stronger comb lines near the DWs because ofer the low 2nd-order dispersion.

\begin{figure*}[t]
\centering
\includegraphics[width=1\linewidth]{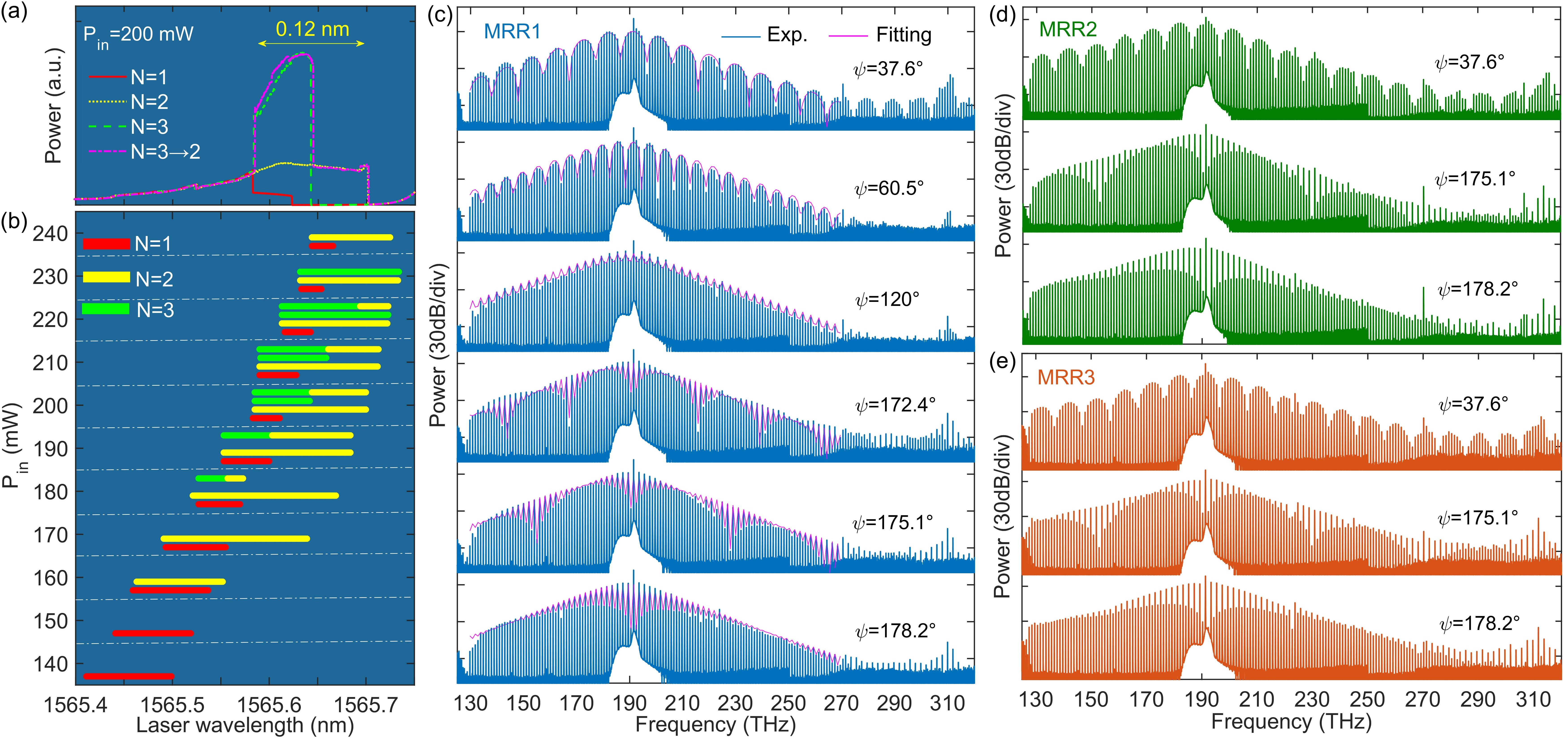}
\caption{\textbf{(a)}-\textbf{(c)} Soliton results of MRR1. \textbf{(a)} Transmitted power excluding pump when \textit{P}$_{in}$= 200 mW. \textbf{(b)} The soliton operation wavelength range and corresponding soliton numbers at different \textit{P}$_{in}$. \textbf{(c)} Typical TSM spectra and the fitting. TSM spectra obtained from \textbf{(d)} MRR2 and \textbf{(e)} MRR3 when \textit {P}$_{in}$ is around 220 and 320 mW, respectively.}
\label{fig4}
\end{figure*}

\section{Experimental results}\label{sec:resul}
\noindent\textbf{Octave-spanning single-soliton and soliton crystals.} In this section, we present the octave-spanning soliton results using the first 1.68-$\mu$m-wide device MRR1 as discussed in Fig. \ref{fig2}. All transmission curves and optical spectra presented in this work are obtained with a forward pump tuning speed of 1 nm/s, which allows for adiabatic tuning within the cavity. The experimental setup is similar to the one reported in \cite{weng2021directly} and can be found in \textbf {supplementary material}. Figure \ref{fig3}(a) shows the simultaneously measured transmission of all output and pump alone, as well as their difference when \textit{P}$_{in}$=150 mW. The striking steps related to the single-soliton formation are observed, with a width of 0.08 nm ($\sim$10 GHz), which is close to $\Delta\lambda$. By stopping the pump at different wavelengths, we map the spectra evolution of the microcomb in Fig. \ref{fig3}(b), which confirms the aforementioned wide SER conclusively. Two dash lines (i) and (ii) indicate the state of MI comb and single-soliton, whose spectra are plotted in Fig. \ref{fig3}(c)-(i) and -(ii). The spectrum ranges from 136 to 240 THz at MI state but is conspicuously widened to 125-322 THz for the single-soliton. The soliton microcomb covers 1.5-octaves and is state of the art \cite{li2017stably, pfeiffer2017octave,briles2018interlocking}. The dual-DW at the frequency of 132 and 311 THz agree with the simulation results very well. The transition from MI to soliton state can also be verified by the drastic reduction of the RF intensity noise, as shown in the inset of Fig. \ref{fig3}(c). For comparison, we also pump the TE$_{00}$ mode at 1549.1 nm which is far from the auxiliary mode. Only an MI comb ranging from 136 to 240 THz appears as a final state when \textit {P}$_{in}$=150 mW (see \textbf {supplementary material}). These results illustrate that the dual-mode scheme could also decrease the pump power required to reach the soliton state.

The carrier-envelope offset frequency (\textit{f}$_{ceo}$) is an important parameter for microcomb in the applications of metrology and timekeeping, which can be detected via the \textit{f}-2\textit{f} self-referencing technique \cite{del2016phase}. A near-zero \textit{f}$_{ceo}$ is ideal for electronic detection and phase-locking \cite{briles2018interlocking}. Figure \ref{fig3}(d) shows the single-soliton spectrum at low-frequency, where the circles and crosses indicate the 1st-order (i.e., realistic) and 2nd-order comb lines, respectively. The latter has a spacing of half FSR and an intensity increasing trend with the frequency decrease. The \textit{f}$_{ceo}$ can be calculated via
\begin{eqnarray}
{f_{ceo}}=2&&\times(f_n-f_{2n}/2)
\end{eqnarray}
\noindent where \textit{f}$_{n}$ and \textit{f}$_{2n}$/2 are frequencies of the adjacent 1st-order and 2nd-order comb lines. Consequently, an \textit{f}$_{ceo}$ of $\sim$200 GHz is extracted. Different from the conventional \textit{f}-2\textit{f} beatnote detection via frequency doubling, our calculation accuracy is only at the level of 1 GHz limited by the OSA resolution. Nevertheless, this simple measurement could help to achieve a low \textit{f}$_{ceo}$ by optimizing the microring dimension further.

The soliton crystal (SC) is an extraordinary state with regularly distributed soliton pulses and enhanced comb line power spaced by multiples of the cavity FSR \cite{cole2017soliton, wang2018robust}. For example, \textit{N}-SC exhibits comb lines separated by \textit{N}×FSR. Such SCs are typically formed in the presence of avoided mode crossing \cite{karpov2019dynamics, weng2021near}. In our scheme, a weak mode coupling occurs between the two neighboring resonances (see \textbf {supplementary material}) and result in the observation of 2-SC and 3-SC when adjusting the power slightly, as shown in Fig. \ref{fig3}(c)-(iii) and -(iv), respectively. Both spectra exceed an octave-spanning range (127-270 THz) and exhibit stronger comb lines near the pump. To our knowledge, this is the first report of octave-spanning dissipative SCs centered on the C-band. As regards 3-SC, there are 8 and 15 comb lines with powers greater than 1 mW and 100 $\mu$W, respectively. The in-waveguide comb powers of the single-soliton, 2-SC, and 3-SC are estimated to be 11.5, 22.8, and 31 mW, corresponding to a conversion efficiency (CE) of 7.7$\%$, 11.4$\%$, and 13.5$\%$. Compared with conventional single DKSs (a few percent CE) \cite{bao2014nonlinear}, the CE of SC is greatly enhanced and we believe it can be further improved by refining the external coupling rate \cite{jang2021conversion}. Figure \ref{fig3}(e) shows the experimental pulse traces carried out by an autocorrelator, where the periods of single-soliton, 2-SC and 3-SC are $\sim$1, $\sim$0.5 and $\sim$0.33 ps, inversely proportional to the \textit{f}$_{rep}$ of $\sim$1, $\sim$2 and $\sim$3 THz. With sech-squared fitting, the pulse width of single-soliton is deduced to be 35 fs from the autocorrelator trace, while an 18 fs width will be resulted from the spectrum by assuming a transform-limited pulse. This discrepancy can be mainly attributed to the phase variation across the pulse spectrum, which can be more precisely determined through a characterization technique like frequency resolved optical gating (FROG). In the \textbf {supplementary material}, we also present the 3-SC and 4-SC from other dual-mode devices.

\begin{figure*}[t]
\centering
\includegraphics[width=1\linewidth]{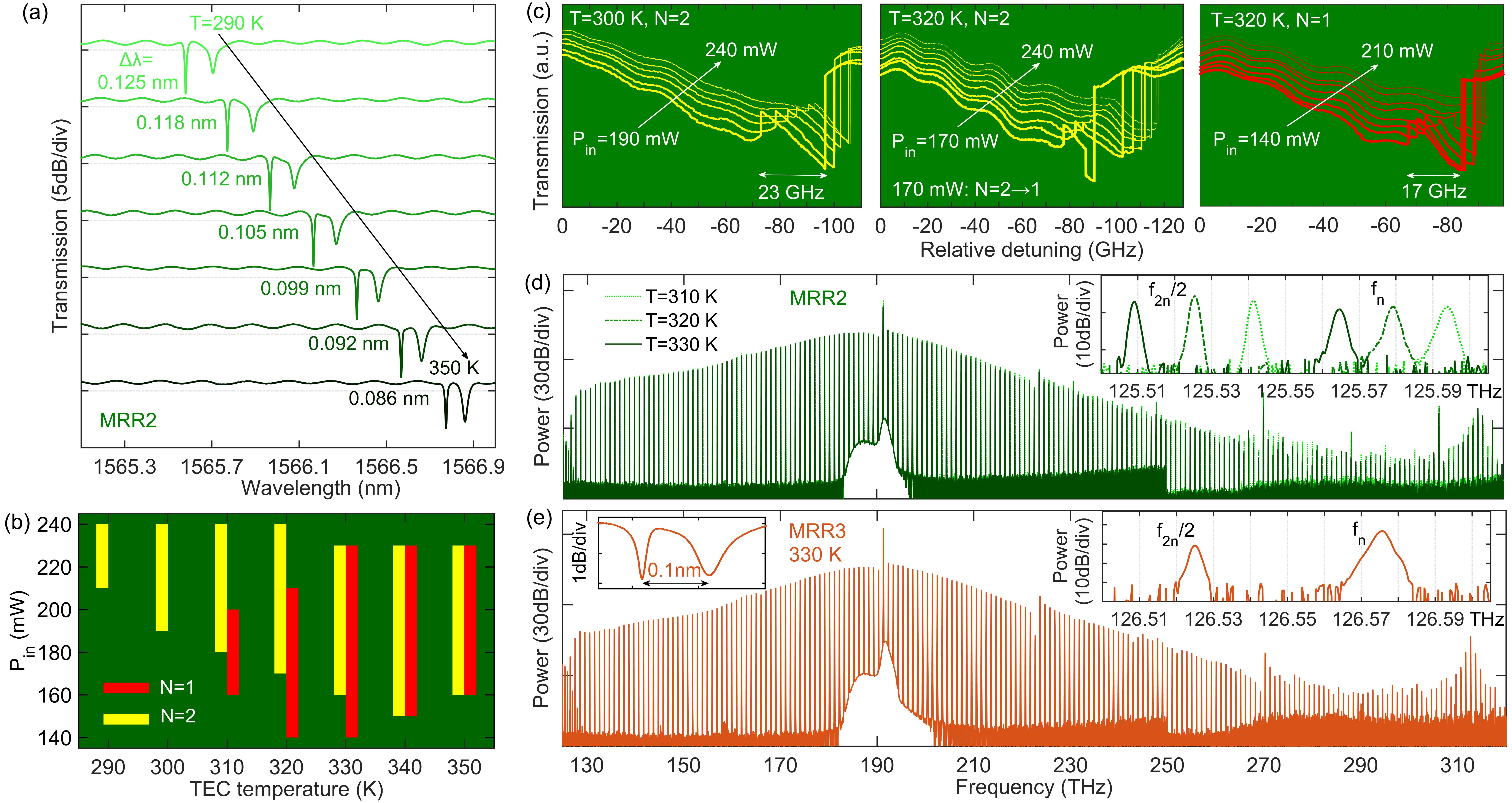}
\caption{\textbf{(a)}-\textbf{(d)} Measurement results with MRR2. \textbf{(a)} Dial-mode transmission spectra at different temperatures. \textbf{(b)} The soliton behavior dependence on the pump power and temperature. \textbf{(c)} Transmission spectra at high pump power with available soliton steps. \textbf{(d)} Single-soliton spectra obtained at 310, 320, and 330 K. \textbf{(e)} Single-soliton spectra achieved with MRR3 when \textit {T}=330 K and \textit {P}$_{in}$=290 mW. Dual-mode transmission (left inset) and two adjacent lines at low-frequency (right inset).}
\label{fig5}
\end{figure*}

\noindent\textbf{Versatile two-soliton microcombs with 1P2R-1P.} 
By repeatedly scanning the laser over the dual-mode in MRR1 at 200 mW, four typical transmission spectra (excluding pump) are observed and recorded, as shown in Fig. \ref{fig4}(a). Besides the single-soliton, we also obtain the multi-solitons with soliton number (\textit{N}) of 2 and 3 as well as the switching from \textit{N}=3 to \textit{N}=2. Particularly, the two-soliton microcomb (TSM) has an SER of 0.12 nm ($\sim$15 GHz). We clarify that these curves are not real comb power considering that some of pump power, even if lower than the parametric threshold, will be absorbed by the auxiliary TE$_{10}$ mode. Despite this, the soliton number is easily identified through the OSA and power meter measurements. Figure \ref{fig4}(b) is a diagram showing the soliton number and effective detuning range dependence on \textit{P}$_{in}$, which indicates that the single-soliton can be definitely generated at 140 and 150 mW. As \textit{P}$_{in}$ increases, both the access probability and SER of single-soliton are decreasing, while the multi-solitons appear with higher possibility. The TSM attained at 180 mW has a 0.15-nm-wide ($\sim$18.8 GHz) SER, near twice that of $\Delta\lambda$. The 1P2R-1P scheme is therefore shown to allow the creation of two-soliton states with ease. 

Figure \ref{fig4}(c) shows several spectra of the TSMs when \textit {P}$_{in}$ is around 200 mW. These soliton combs can be reproduced easily and sustained for long periods without noise. The relatively azimuthal angles of 37.6°, 60.5°, 120°, 172.4°, 175.1°, and 178.2° are retrieved by fitting the spectral envelope with
\begin{eqnarray}
S^{(2)}(\mu)=S^{(1)}(\mu)\times(2+2\times cos(\mu\psi))
\end{eqnarray}
 \noindent where $\psi$ is the relative azimuthal angle between the two pulses, $\mu$ is the comb mode index relative to the pump position, and \textit{S}$^{(1)}$($\mu$) is the spectral of a single-soliton following a sech-squared shape fitted from the experimental data \cite{doi:10.1126/science.aad4811}. The TSMs with $\psi$ of 37.6°, 175.1° and 178.2° are reproducible in the other two 1.68-$\mu$m-wide devices (i.e., MRR2 and MRR3), as shown in Figs. \ref{fig4}(d) and \ref{fig4}(e), respectively. All resonators are over-coupled at the pump wavelength, while the MRR1, MRR2, and MRR3 have a coupling gap of 650, 650, and 550 nm, respectively. Thus for MRR3, the pump power to access solitons is higher and the powers of individual lines are stronger at both high and low frequencies. These TSMs have a generally higher CE compared with the single-soliton, especially for the $\psi$=178.2° case, which has a CE beyond 10$\%$. Such diverse soliton states with improved CE are of interest in applications such as optical arbitrary waveform generation \cite{jiang2007optical} and microwave photonic filters (with larger resonators) \cite{xu2019high}. The results of three-soliton microcombs are shown in the \textbf {supplementary material}.
 \\
\noindent\textbf{Thermal control for separation tuning and single-soliton generation.} 
All above results are obtained by maintaining the substrate temperature at 290 K (17 °C) with the aid of a thermoelectric cooler (TEC). However, only multi-solitons can be triggered in MRR2 and MRR3, which possess a relatively large $\Delta\lambda$ of 0.125 and 0.127 nm. Next, we will show the effective control of the mode separation and soliton state by changing TEC temperature \textit{T}. Figure \ref{fig5}(a) draws the dual-mode transmission spectra in MRR2 at different temperatures. As the temperature increases, a thermally induced redshift of the resonant wavelengths are observed with a d$\lambda$/d\textit{T} of $\sim$0.02 nm/K, corresponding to a thermo-optic coefficient of $\sim$ 2.3$\times$10$^{-5}$/K, which is consistent with the result reported in \cite{arbabi2013measurements}. The $\Delta\lambda$ declines from 0.125 to 0.086 nm when \textit{T} increases from 290 to 350 K, indicating that the resonant wavelength of the TE$_{00}$ mode is more sensitive to the temperature variation. We also note that the coupling between the two modes is strengthened when they approach each other, leading to the increase in the extinction ratio of the TE$_{10}$ mode. Figure \ref{fig5}(b) depicts the relation between on-chip power and soliton number at various temperatures. It can be seen that only TSMs can be reached at the temperature of 290 and 300 K, while the single-solitons arise when \textit{T}$\geq$310 K. When 310$\leq$\textit{T}$\leq$330 K, low power can trigger single-soliton only, while conversely, TSMs tend to be formed at high power, which is similar to the trend shown in Fig. \ref{fig4}(b).

\begin{table*}
  \centering
  \newcommand{\tabincell}[2]{\begin{tabular}{@{}#1@{}}#2\end{tabular}}
	\caption{Comparison of octave-spanning DKSs.} \label{table1}
	\setlength{\tabcolsep}{1mm}{}
  \begin{tabular}{lcccccccc}
  
  	  \toprule
    \tabincell{c}{Material} & 
    \tabincell{c}{\textit{Q}-factor\\(million)} & 
    \tabincell{c}{On-chip \\power (mW)}  & \tabincell{c}{Spectral range\\(THz)}  & \tabincell{c}{\textit{f}$_{rep}$\\(THz)} & 
    \tabincell{c}{SER\\(GHz)} & 
     Accessing method \\ 
    
		Si$_3$N$_4$\cite{li2017stably} & 2 (\textit{Q}$_{int}$) & 120$\pm$15 & 129-275 & $\sim$1 & $\sim$1.5 & \tabincell{c}{Adiabatic pump sweeping (-100 GHz/s)\\ with 1P2R-2P scheme}  \\
		Si$_3$N$_4$\cite{pfeiffer2017octave} & $\sim$1(\textit{Q}$_{load}$) & 455 & 130-280 & $\sim$1 & - & \tabincell{c}{Forward sweeping and backward tuning}  \\ 
		Si$_3$N$_4$\cite{briles2018interlocking} & - & 200 & 130-310 & $\sim$1 & - & Fast pump sweeping with frequency shifter \\
		Si$_3$N$_4$\cite{briles2021hybrid} & 2.7(\textit{Q}$_{int}$) & 40 & 140-280 & $\sim$1 & - & \tabincell{c}{Self-injection locking} \\
		AlN\cite{weng2021directly} & 1.4(\textit{Q}$_{int}$) & $\sim$335 & 130-273 & $\sim$0.37 & $\sim$10.4 & \tabincell{c}{Adiabatic pump sweeping (-125 GHz/s)\\ with 1P2R-1P scheme}  \\
		AlN\cite{liu2021aluminum} & 1.6(\textit{Q}$_{int}$) & $\sim$390 & 130-295 & $\sim$0.43 & - & \tabincell{c}{Fast pump sweeping with \\ single-sideband modulator}  \\
		LiNbO$_3$\cite{he2021octave} & 1.15(\textit{Q}$_{load}$) & $\sim$600 & 125-268 & $\sim$0.2 & $\sim$0.2 & \tabincell{c}{Self-start (photorefractive effect)}  \\\tabincell{c}{\textbf{Si$_3$N$_4$}\\\textbf {(this work)}}	& $\sim$1.1(\textit{Q}$_{int}$) & \tabincell{c}{140\\ (200, 230)} & \tabincell{c}{125-320\\ (127-270)} & \tabincell{c}{$\sim$1\\ ($\sim$2, $\sim$3)} & $\sim$17 & \tabincell{c}{Adiabatic pump sweeping (-125 GHz/s)\\ with 1P2R-1P scheme}  \\
  \end{tabular}
\end{table*}

Figure \ref{fig5}(c) shows examples of transmission spectra measured at high power, where the relevant soliton number is labelled at the top. Specifically, the TSM with a notable SER of $\sim$23 GHz is observed when \textit{T}=300 K and \textit{P}$_{in}$=190 mW. At 320 K, a soliton switching from \textit{N}=2 to \textit{N}=1 is observed when \textit{P}$_{in}$=170 mW. The deterministic access to single-soliton state is also realized at 320 K, accompanied by a maximum SER of $\sim$17 GHz, slightly wider than the 0.105-nm-wide $\Delta\lambda$ ($\sim$13 GHz). The single-soliton spectra acquired at 310, 320, and 330 K are plotted in Fig. \ref{fig5}(d) with pump wavelengths of 1566.550, 1566.750, and 1566.950 nm, respectively. The spectra have similar profiles and range from 125 to 320 THz, well beyond an octave span. As with the results originating from MRR1, dual-DW at the frequency of 130 and 312 THz are observed. The inset exhibits the two adjacent 1st-order and 2nd-order comb lines near 125.5 THz. At 310, 320, and 330 K, the \textit{f}$_{ceo}$ is calculated to be $\sim$105, $\sim$108 and $\sim$109 GHz, respectively, which is about half of the \textit{f}$_{ceo}$ shown in Fig. \ref{fig3}(d) from MRR1. 

By setting the temperature of MRR3 at 330 K and tuning the pump wavelength to 1566.885 nm, the single-soliton can be stably accessed [see Fig. \ref{fig5}(e)], which has a similar profile as Fig. \ref{fig3}(c)-(ii) (MRR1) and Fig. \ref{fig5}(d) (MRR2). In this case, the $\Delta\lambda$, SER and \textit{f}$_{ceo}$ are $\sim$0.1 nm ($\sim$12.5 GHz), $\sim$11 GHz and $\sim$100 GHz, respectively. It should be mentioned that the MRR2 and MRR3, with almost identical mode separation and \textit{f}$_{ceo}$, are in the same chip, which strongly suggests the fabrication uniformity. These results indicate that temperature control will be crucial for the deterministic creation of single-solitons.

\section{Discussion}
\noindent The results demonstrate that the 1P2R-1P scheme is applicable to our 23-$\mu$m-radius Si$_3$N$_4$ MRRs with the proper mode separation (e.g., 10-13 GHz). The SER of single-soliton is generally equivalent to $\Delta\lambda$, while the window of multi-soliton is up to almost twice $\Delta\lambda$. Table \ref{table1} compares the reported octave-spanning DKSs realized with various platforms. Clearly, the SER is much expanded with an auxiliary resonance in our 1P2R-1P scheme. More importantly, the proposed strategy enables the access to soliton state via straightforward pump frequency control instead of rapid frequency scan or complicated control. The present octave-spanning single-solitons are generated by the Si$_3$N$_4$ MRRs with \textit{Q} of $\sim$1.1 million, but the ongoing experiments suggest that a 2.7 million \textit{Q} will reduce the required on-chip power to \textless 40 mW \cite{briles2021hybrid}, which paves a way towards a miniaturized soliton system that is integrated with a laser diode. We also demonstrate the octave-spanning soliton crystals generation.

We believe that the proposed 1P2R-1P approach for deterministic access to DKS could have profound significance on the microcomb field if a design could be reproduced in fabrication with a reasonable yield. Some solutions can be adopted to further control the mode separation and improve the yield. First, according to Fig. \ref{fig2}, a fine (a few nm) ring dimension scan is necessary when designing the layout to ensure the accuracy of relative variation in fabrication. Considering the reliability, uniformity, and ability to fabricate a high density of MRRs in the commercial foundry, a reasonable variation in MRR dimensions is possible to provide more samples featuring both the desired dual-mode and low \textit{f}$_{ceo}$. Second, post-fabrication such as etching \cite{moille2021tailoring} can be used to tune the resonance characteristics and the mode spacing. Finally, as investigated with MRR2 and MRR3, temperature control can effectively modify the mode separation and change the soliton state. The control can be implemented by a substrate TEC or surface microheaters, which has been demonstrated for the thermal tuning of \textit{f}$_{rep}$ and \textit{f}$_{ceo}$ \cite{xue2016thermal}. In practice, we can also tune pump wavelength or change the laser source if the real dual-mode region deviates from the designed position.

\section{Conclusions}
\noindent In summary, we demonstrate the accessing of octave-spanning single-soliton, soliton crystals, and multi-solitons in the dual-mode microresonators via simply slow pump tuning. In addition to rich soliton states, the conventional inaccessible soliton step is stabilized now accompanied by an expanded detuning range. Compared with the results achieved by the 2P2R method using an independent auxiliary laser \cite{zhou2019soliton}, the demonstrated SER of 17 GHz here has been significantly enhanced by two orders of magnitude. Such a broad soliton existence window will greatly enhance the potential for microcomb use in applications such as parallel FMCW LiDAR \cite{riemensberger2020massively}. The soliton behavior dependence on the pump power and temperature is also explored. 

The proposed straightforward and low-cost soliton generation system 1P2R-1P is universally feasible for appropriately designed microresonators. We have demonstrated this first in an AlN MRR and expanded on this in the present study. It should also be possible across other platforms for soliton microcomb generation in LiNbO$_3$ and SiC as examples. It is foreseeable that, by using a passive dual-mode microresonator with an upgraded \textit{Q} such as $\sim$5×10$^6$ \cite{ji2021methods}, the photonic integrated octave-spanning coherent microcomb source will be delivered soon, which is driven by a laser diode with a power of about 100 mW but without amplifier or optical feedback. We also note that a recent work \cite{lei2022thermal} shows the positive effects of 1P2R-2P in improving the timing jitter and effective linewidth of the soliton microcomb lines. We believe such a simple system and the versatile soliton states will not only accelerate the achievement of commercial, portable and affordable soliton microcomb sources but also contribute to the extension of their applications.
\\

\noindent\textbf{Supplementary material} See the \textbf {supplementary material} for additional information.
\\

\noindent\textbf{Acknowledgements} This project is supported by the Science Foundation Ireland (Grant No. 17/NSFC/4918) and the National Natural Science Foundation of China (Grant No. 61861136001).
\\

\noindent\textbf{Competing interests} The authors have no conflicts of interest to disclose.
\\

\noindent\textbf{Data availability} The data in my manuscript can be obtained from the corresponding author.

% Please use pisikabst.bst. You may your own *.bib file.
\bibliographystyle{pisikabst}
\bibliography{manuscript}

\pagebreak
\widetext
\begin{center}
\textbf{\large Supplementary Material: Dual-mode microresonators as straightforward access to \\octave-spanning dissipative Kerr solitons}
\end{center}

\setcounter{equation}{0}
\setcounter{figure}{0}
\setcounter{table}{0}
\setcounter{section}{0}
\setcounter{page}{1}
\makeatletter
\renewcommand{\theequation}{S\arabic{equation}}
\renewcommand{\thefigure}{S\arabic{figure}}
\renewcommand{\bibnumfmt}[1]{[S#1]}
\renewcommand{\citenumfont}[1]{S#1}
%%%%%%%%%% Prefix a "S" to all equations, figures, tables and reset the counter %%%%%%%%%%

\section{Experiment setup}

Figure S1 shows the experimental setup for soliton microcomb generation and characterization. The light source used in the experiment is a tunable laser ranging from 1480 to 1640 nm (Santec TSL-710). An erbium-doped fiber amplifier (EDFA) is adopted to amplify the laser power for pumping the resonator through a lensed fiber with a spot size of 2.5 $\mu$m. A fiber polarization controller (FPC) is utilized to adjust the polarization. The generated soliton microcombs are collected with another lensed fiber for characterization. The optical spectra are recorded by two optical spectrum analyzers (1200-2400 nm and 600-1700 nm range). By injecting the output into an electrical spectrum analyzer (ESA) after a photodiode (PD), the radio-frequency (RF) intensity noise of the combs can be detected. The autocorrelation measurement shown in Fig. \ref{fig3}(e) is carried out by the autocorrelator (AC).  For the transmission measurement, the output is divided into two parts, one is for all-transmission and the other one is for the pump-transmission only after a band-pass filter. By calculating the difference between the two branches, we can easily identify the soliton steps in the transmitted curves, as shown in Figs. \ref{fig3}(a) and \ref{fig4}(a) in the main text.

\begin{figure}[htbp]
\centering
\includegraphics[width=0.72\linewidth]{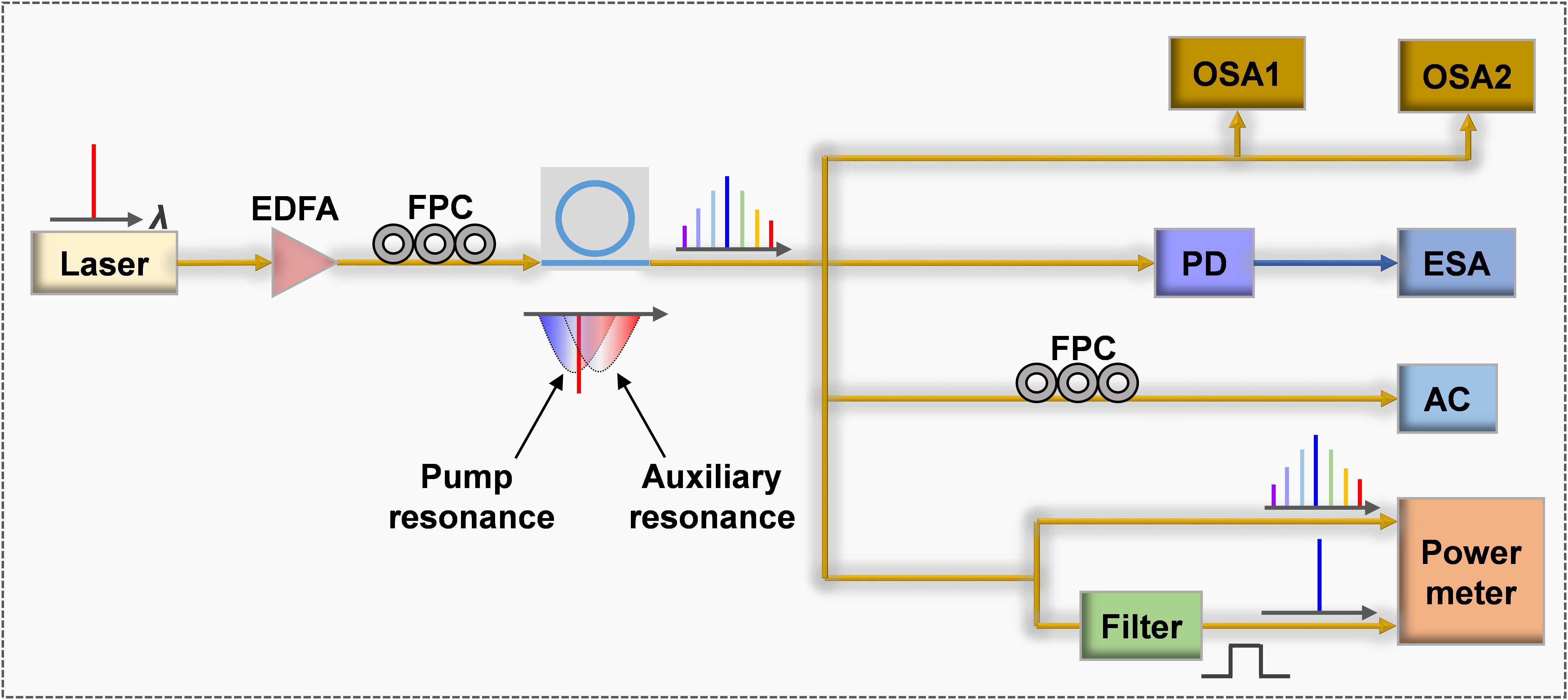}
\caption{Experimental setup used for dissipative Kerr solitons (DKSs) generation. FPC, fiber polarization controller; TEC, thermoelectric cooler; EDFA, erbium-doped fiber amplifier; OSA, optical spectrum analyzer; OSA1, 600–1700 nm; OSA2, 1200–2400 nm; FBG, fiber Bragg grating; PD, photodiode; ESA, electrical spectrum analyzer; AC, autocorrelator.}
\label{figs1}
\end{figure}

\section{Mode coupling in MRR1 and the generated MI comb}

The transmission spectrum of MRR1 measured in a wide range is shown in Fig. S2(a), where the TE$_{00}$ and TE$_{10}$ modes are marked by circles and crosses, respectively. The dual-mode near 1565 nm used for the soliton generation is labelled with a relative mode number $\mu$=0. Figure S2(b) plots the experimental loaded \textit Q factors, where the values of TE$_{00}$ modes have an abrupt change between $\mu$=0 and $\mu$=-1 due to the mode coupling. The extinction ratio of TE$_{00}$ mode also changes significantly at $\mu$=-1. Figure S2(c) shows the free spectra ranges (FSRs) of the two mode families. Clearly, the fundamental and first-order modes have a statistical FSR of $\sim$1012 and $\sim$979 GHz, respectively. However, because of the mode coupling, the obvious FSR changes are observed for the TE$_{00}$ mode at $\mu$=0 as well as TE$_{10}$ modes at $\mu$=-1 and -2. Such doublet resonances with avoided mode crossing are pretty general especially in the resonators with relatively small FSR [1-3]. It has also been exploited for dispersive wave formation [4] and soliton crystal (SC) generation [5].

\begin{figure}[htbp]
\centering
\includegraphics[width=0.95\linewidth]{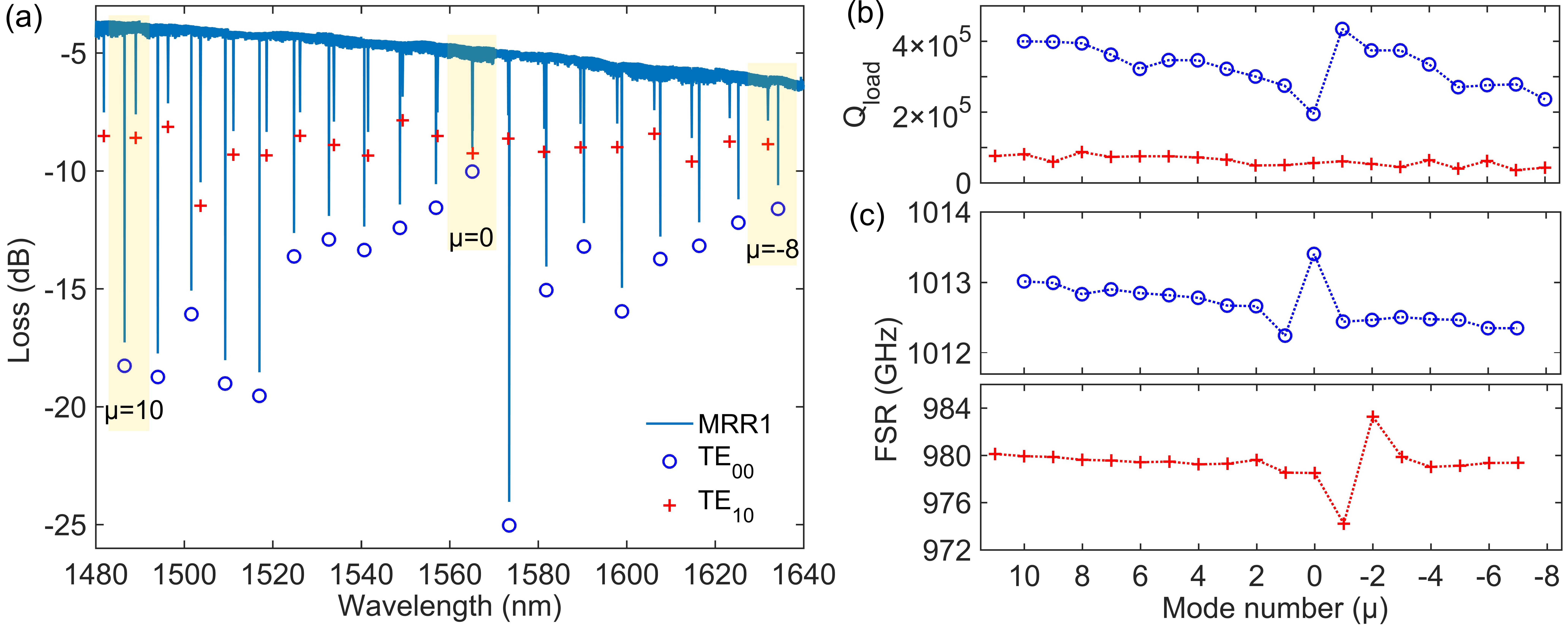}
\caption{\textbf{(a)} Transmission spectrum of MRR1, where the TE$_{00}$ and TE$_{10}$ modes are marked by circles and crosses, respectively. The resonances at relative mode number $\mu$ = 0 have a minimum separation of 0.084 nm. Experimental \textbf{(b)} \textit Q factors and \textbf{(c)} FSRs of the TE modes. }
\label{figs2}
\end{figure}

\begin{figure}[htbp]
\centering
\includegraphics[width=0.9\linewidth]{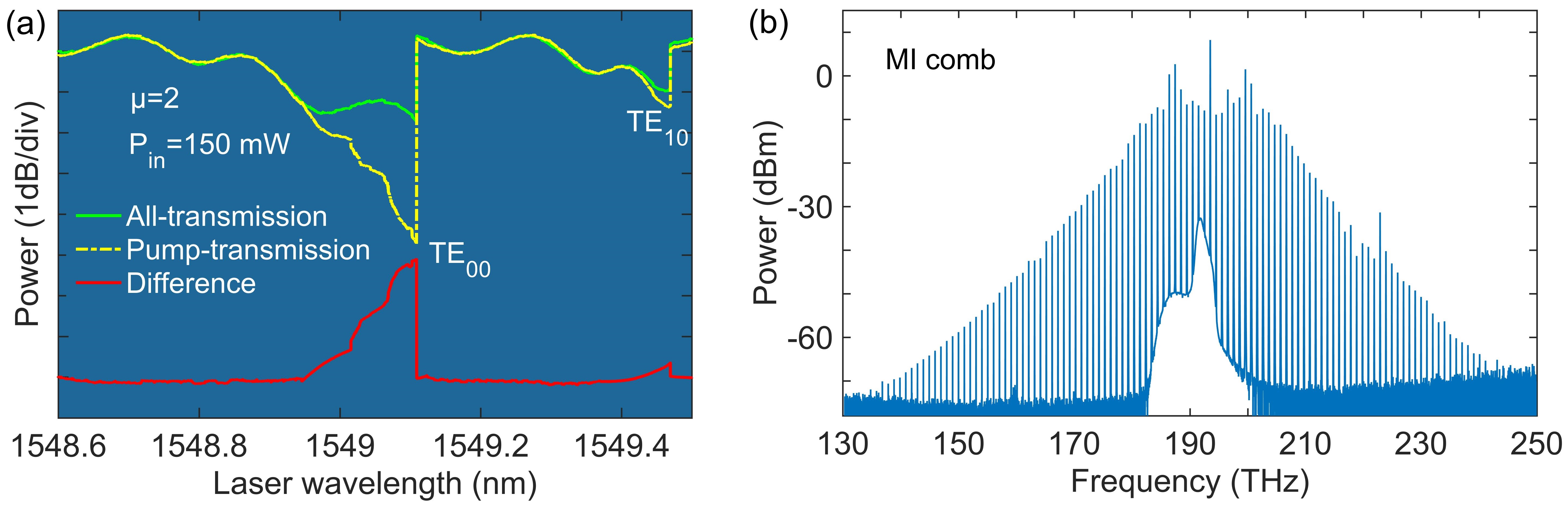}
\caption{\textbf{(a)} The collected output power of all-transmission, pump-transmission, and their difference when pumping the TE$_{00}$ mode at $\mu$=2. \textbf{(b)} Measured spectrum of MI comb.}
\label{figs3}
\end{figure}

For comparison, we also pump the TE$_{00}$ mode at $\mu$=-2, which has a separation of 0.62 nm with the adjacent TE$_{10}$ mode. Figure S3(a) shows the measured transmission spectra at an on-chip pump power (\textit P$_{in}$) of 150 mW. There is no overlap between the two resonances, indicating that the modes will be pumped independently. By sweeping the pump wavelength to 1549.11 nm, just before dropping out of the TE$_{00}$ resonance, we can successfully access the modulation instability (MI) comb. As shown in Fig. S3(b), the MI comb ranges from 135 to 242 THz, below an octave span. It has a similar envelope like the one obtained from the fundamental mode at $\mu$=0 [see Fig.\ref{fig3}(c)-(i)]. Thus the dual-mode scheme could decrease the required power to reach the soliton state.

\section{Three-soliton state and soliton crystals from various MRRs}
In this section, we will introduce the other two 1.68-$\mu$m-wide dual-mode MRRs firstly and then present the versatile soliton states. Figures S4(a) and S4(c) show the transmission spectra of the dual-mode in MRR4 and MRR5, which have a separation of 0.067 and 0.019 nm, respectively. The transmitted power curves at 200 mW excluding the pump of MRR4 are plotted in Fig. S4(b), where the visible steps correspond to the formation of two-soliton (0.02-nm-wide) and three-soliton (0.11-nm-wide) states. For MRR5, the extinction ratio of the TE$_{10}$ resonance is even higher than the TE$_{00}$ resonance, suggesting a strong mode coupling between the two modes. 3- and 4-SC are realized with MRR5 using the slow pump sweeping technique. There is no single-soliton observation for these two resonators when increasing the TEC temperature, which we attribute to the relatively small mode separations. Nevertheless, deterministic generation of the multi-solitons with pronounced soliton existence window manifests the reliability and practicality of our approach. 

Figure S4(d) shows the octave-spanning three-soliton combs generated with MRR1, MRR3, and MRR4 under the pump power of 210, 320, and 200 mW, respectively. Similar three-soliton states are reproducible among different chips. In addition, the three-solitons obtained from MRR3 have stronger intensities due to the narrower coupling gap between resonator and waveguide. Figure S4(e) summarizes the spectra of SCs. The 3-SC spectra originating from MRR4 exceeds an octave-spanning (130-270 THz) with an estimated conversion efficiency of 12.8$\%$. However, because of the smaller mode separation and enhanced mode coupling in MRR5, the 3-SC and 4-SC have a relatively narrow bandwidth (140-230 THz). 

Overall, it is easier to access the single-soliton state in the dual-mode resonators (MRR1, MRR2, and MRR3) with a relatively large mode separation (e.g., 10-13 GHz). However, we perceive that the mode separation is a critical but not the only factor that affects the soliton behavior. More intensive work needs to be implemented in the future to sort out the influence of mode coupling on thermal compensation and soliton behavior.
\\
\begin{figure}[t]
\centering
\includegraphics[width=0.96\linewidth]{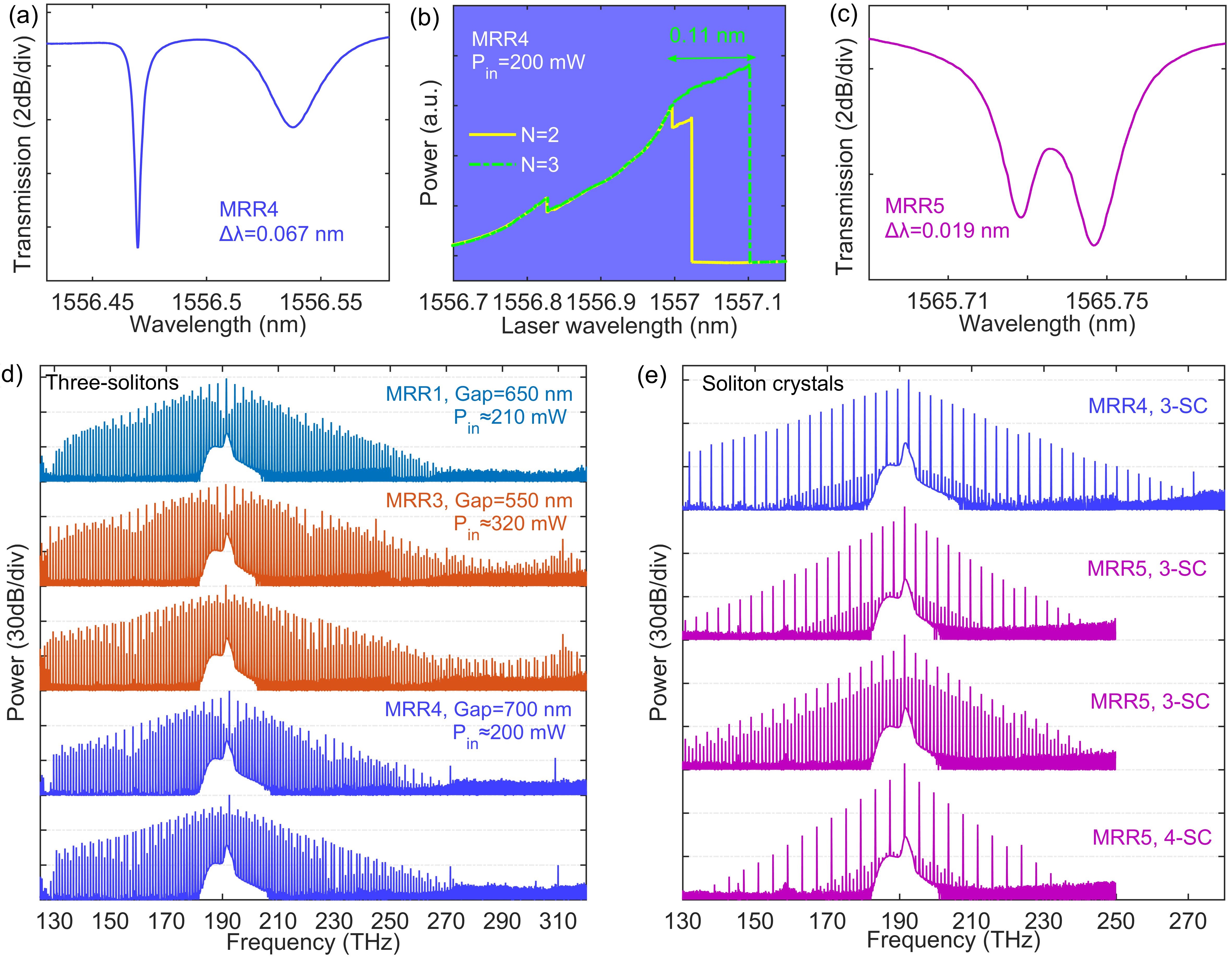}
\caption{Transmission spectrum for the dual-mode in \textbf{(a)} MRR4 and \textbf{(c)} MRR5. \textbf{(b)} Measured transmitted power curves excluding the pump in MRR4 when \textit P$_{in}$=200 mW. Experimental spectra of \textbf{(d)} three-solitons and \textbf{(e)} soliton crystals from different microresonators.}
\label{figs4}
\end{figure}

\noindent\textbf{References}

\noindent[S1] S. Ramelow, A. Farsi, S. Clemmen, J. S. Levy, A. R. Johnson, Y. Okawachi, M. R. Lamont, M. Lipson, and A. L. Gaeta, "Strong polarization mode coupling in microresonators," Optics Letters 39, 5134-5137 (2014).
\\
\noindent[S2] Y. Liu, Y. Xuan, X. Xue, P.-H. Wang, S. Chen, A. J. Metcalf, J. Wang, D. E. Leaird, M. Qi, and A. M. Weiner, "Investigation of mode coupling in normal-dispersion silicon nitride microresonators for Kerr frequency comb generation," Optica 1, 137-144 (2014).
\\
\noindent[S3] E. Nazemosadat, A. Fülöp, Ó. B. Helgason, P.-H. Wang, Y. Xuan, D. E. Leaird, M. Qi, E. Silvestre, A. M. Weiner, and V. Torres-Company, "Switching dynamics of dark-pulse Kerr frequency comb states in optical microresonators," Physical Review A 103, 013513 (2021).
\\
\noindent[S4] Q.-F. Yang, X. Yi, K. Y. Yang, and K. Vahala, "Spatial-mode-interaction-induced dispersive waves and their active tuning in microresonators," Optica 3, 1132-1135 (2016).
\\
\noindent[S5] M. Karpov, M. H. Pfeiffer, H. Guo, W. Weng, J. Liu, and T. J. Kippenberg, "Dynamics of soliton crystals in optical microresonators," Nature Physics 15, 1071-1077 (2019).

\end{document}